\documentclass[aps]{revtex4}%
\usepackage{amsfonts}
\usepackage{amsmath}
\usepackage{amssymb}
\usepackage{graphicx}%
\begin{document}
\title[Generalized Einstein relation]{Generalized Einstein relation in conductors with a \ large built-in field}
\author{Sergey A. Ktitorov}
\affiliation{Ioffe Institute, Polytechnicheskaya str. 26,  \ 194021 St. Petersburg, Russia
and State Electrotechnical University, Prof. Popov str. 5, 197376 St.
Ptersburg, Russia}

\begin{abstract}
Generalized Einstein relation between the mobility and diffusion in conductors
with a large built-in field near the thermodynamic equilibrium has been derived.

\end{abstract}
\maketitle









\qquad The mobility and diffusion coefficients depend on the particle energy.
When the system under consideration is in the inequilibrium state and the
current does not vanish, the electons are hot and their energy is determined
by the balance of power, which electrons acquire from the electric field and
transfere to the thermostat. Mobility is determined by the mean kinetic energy
of electrons and within the framework of the quasi-hydrodynamic approximation
can be considered as a function of the local electric field, if
inhomogeneities are spread along the current lines \cite{shura} (this is the
case for N-shape current-voltage characteristics). Attempts to generalize the
Einstein relation are directed mainly to analysis of strongly inequilibrium
state (see \cite{strong}, for instance). In the opposite case of S-shape
charaterisitics the mobility can be considered as a function of the local
current. We encounter quite different situation, when the system is near the
thermodynamic equilibrium, but the built-in electric field is strong enough so
that the mobility depends mainly on the potential energy: $\mu=\mu\left(
\phi\right)  .$ Therefore, the diffusion coefficient depends on the potential
as well: $D=D\left(  \phi\right)  .$ We consider here a one-dimensional
distribution of the electric field and the electronic density. The total
current reads:%
\begin{equation}
j=-en\left(  x\right)  \mu\left[  \phi\left(  x\right)  \right]  \frac{d\phi
}{dx}-e\frac{d}{dx}\left[  D\left(  \phi\right)  n\left(  x\right)  \right]
,\text{ \ \ , }E=-d\phi/dx,\text{ \ \ }\frac{dj}{dx}=0. \label{current}%
\end{equation}

We have in the thermodynamic equilibrium:%
\begin{equation}
j=0\text{, \ \ }n\left(  x\right)  =n_{0}e^{-\frac{e\phi}{T}}.
\label{equilibrium}%
\end{equation}

Carrying out differetiation in (\ref{current}) we have%
\begin{equation}
j=-en\left(  x\right)  \mu\left[  \phi\left(  x\right)  \right]  \frac{d\phi
}{dx}-e\frac{dD}{d\phi}n\left(  x\right)  \frac{d\phi}{dx}-eD\left(
\phi\right)  \frac{dn}{d\phi}\frac{d\phi}{dx}, \label{current2}%
\end{equation}

where $\frac{dn}{d\phi}=-\frac{e}{T}n.$ Thus, we obtain the differential
equation for the diffusion coefficient:%
\begin{equation}
\frac{dD}{d\phi}+\frac{d\log n}{d\phi}D+\mu\left(  \phi\right)  =0
\label{diffusion}%
\end{equation}

or%

\[
\frac{dD}{d\phi}-\frac{e}{T}D+\mu\left(  \phi\right)  =0
\]

Let us denote $\mu\left(  \phi=0\right)  \equiv\mu_{0}.$ Then we can introduce
the dimensionless variables:%
\begin{equation}
\widetilde{\mu}=\frac{\mu}{\mu_{0}},\text{ \ \ }\psi=\frac{e\phi}{T},\text{
\ \ }\widetilde{D}=\frac{eD}{\mu_{0}T}. \label{dimensionless}%
\end{equation}

We obtain the following differential equation for the dimensionless diffusion
coefficient:%
\begin{equation}
\frac{d\widetilde{D}}{d\psi}-\widetilde{D}+\widetilde{\mu}\left(  \psi\right)
=0 \label{dimlessdiffus}%
\end{equation}

Formulae (\ref{dimensionless}) show that the dimensional factor in the
diffusion coefficient has the standart form \ $D/\widetilde{D}=\mu_{0}T/e.$
However, the relation between the diffusion coefficient and the mobility is
non-local, as it can be seen from the equation (\ref{dimlessdiffus}). The
solution of this equation with the initial condition $\widetilde{D}\left(
0\right)  =$ $\widetilde{D}_{0}$ reads:%
\begin{equation}
\widetilde{D}\left(  \psi\right)  =\exp\left(  \psi\right)  \left[
\widetilde{D}_{0}-\int_{0}^{\psi}dz\widetilde{\mu}\left(  z\right)
\exp\left(  -z\right)  \right]  \label{dimlesssolution}%
\end{equation}

Let us consider the trivial case of constant mobility: $\mu=\mu_{0},$ and,
therefore, $\widetilde{\mu}=1,$ $\ \widetilde{D}_{0}=1.$ The solution
(\ref{dimlesssolution}) takes the form:%
\begin{equation}
\widetilde{D}\left(  \psi\right)  =\exp\left(  \psi\right)  \left[  1-\int
_{0}^{\psi}dz\exp\left(  -z\right)  \right]  =1. \label{constant}%
\end{equation}

This calculation confirms consistency of our description.

\qquad Now we consider the case of the exponential law mobility dependence
$\widetilde{\mu}=\exp\left(  a\psi\right)  .$%
\begin{equation}
\widetilde{D}\left(  \psi\right)  =\exp\left(  \psi\right)  -\exp\left(
\psi\right)  \int_{0}^{\psi}dz\exp(\left(  a-1)z\right)  =\frac{a\exp\left(
\psi\right)  }{a-1}-\frac{\exp\left(  a\psi\right)  }{a-1} \label{exp}%
\end{equation}

This expression tends to unity, when $a\longrightarrow0$ or $\psi
\longrightarrow0.$ The diffusion coefficient vanishes identically at $a-1.$

\qquad Let us consider now the case of linear dependence of the mobility on
the potential: $\widetilde{\mu}=1+a\psi.$ Then we have:%
\begin{equation}
\widetilde{D}\left(  \psi\right)  =\exp\left(  \psi\right)  \left[  1-\int
_{0}^{\psi}dz\exp\left(  -z\right)  \right]  =2+\psi-\exp\left(  \psi\right)
. \label{linear}%
\end{equation}

Thus, we see that the diffusion vs mobility relation near the equilibrium is
generically non-local. The conventional Einstein relation $D=\mu T/e$ is valid
in the limit of low built-in potential.

\end{document}